\documentclass[pra,reprint,amsmath,amssymb,aps,superscriptaddress,nofootinbib]{revtex4-1}

\usepackage{amsthm}
\usepackage{algorithm}
\usepackage{algpseudocode}
\usepackage{tabularx}
\usepackage{graphicx}
\usepackage{dcolumn}
\usepackage{bm}
\usepackage{float}
\usepackage{appendix}
\usepackage{comment}
\usepackage[shortlabels]{enumitem}
\usepackage[colorlinks, linkcolor=black,anchorcolor=blue,citecolor=blue,urlcolor=blue]{hyperref}
\usepackage[dvipsnames]{xcolor}
\usepackage{soul}
\usepackage[english]{babel}
\usepackage{blindtext}
\usepackage{todonotes}
\usepackage{url}
\usepackage[most]{tcolorbox}
\usepackage{enumitem} 
\usepackage{tikz}
\usepackage{tikzscale}
\usetikzlibrary{shapes.geometric, arrows}

\usepackage{subfigure} 

\usepackage{subfiles} 

\newtcbox{\button}{on line, 
                    nobeforeafter, 
                    colframe=black, 
                    colback=yellow!50!white, 
                    boxrule=0.2mm, 
                    arc=0pt, 
                    boxsep=0pt, 
                    left=5pt, 
                    right=5pt, 
                    top=2pt, 
                    bottom=2pt, 
                    tcbox raise base,
                    valign=center, 
                    before upper=\strut, 
                    enhanced}

\newtcolorbox{multilinebutton}[1][]{ 
    colframe=black, 
    colback=yellow!50!white, 
    boxrule=0.2mm, 
    arc=0pt, 
    boxsep=0pt, 
    left=5pt, 
    right=0pt, 
    top=2pt, 
    bottom=2pt, 
    notitle, 
    #1
}

\begin{document}

\title{Low-latency control system for feedback experiments with optical tweezer arrays}

\author{Amir~H.~Dadpour}
\affiliation{Institute for Quantum Computing, University of Waterloo, Canada.}
\author{Timur~Khayrullin}
\affiliation{Institute for Quantum Computing, University of Waterloo, Canada.}
\author{Fouad~Afiouni}
\affiliation{Department of Computer Science, American University of Beirut, Lebanon.}
\affiliation{School of Electrical and Computer Engineering, Purdue University, USA.}
\author{Remy~El~Sabeh}
\affiliation{David R. Cheriton School of Computer Science, University of Waterloo, Canada.}
\author{Amer~E.~Mouawad}
\affiliation{Department of Computer Science, American University of Beirut, Lebanon.}
\affiliation{David R. Cheriton School of Computer Science, University of Waterloo, Canada.}
\author{Izzat~El~Hajj}
\affiliation{Department of Computer Science, American University of Beirut, Lebanon.}
\author{Alexandre~Cooper}
\email[]{alexandre.cooper@uwaterloo.ca}
\affiliation{Institute for Quantum Computing, University of Waterloo, Canada.}
\affiliation{Department of Physics and Astronomy, University of Waterloo, Canada.}

\date{\today}

\begin{abstract}
We present and characterize a modular, open-source system to perform feedback control experiments on configurations of atoms and molecules in arrays of optical tweezers. The system features a modular, cost-effective computer architecture with a motherboard and peripheral cards. It supports efficient data transfer to and from graphics processing units (GPUs) using Remote Direct Memory Access (RDMA), leveraging GPU efficiency in matrix multiplication and parallelism, while enabling direct data transfer between devices without involving the CPU.
We first describe the architecture and workflow of the system, detailing its hardware components and software modules. We then evaluate the computational runtime for preparing defect-free chains and grids of atoms using efficient implementations of atom reconfiguration algorithms. Finally, we discuss timing bottlenecks and strategies to reduce latency. Beyond solving reconfiguration problems, the system can readily be used to implement adaptive and feedforward protocols, as well as digital quantum algorithms relying on particle displacement. Our results lay the groundwork for developing low-latency feedback control systems, benchmarking their performance, and advancing scalable quantum hardware.
\end{abstract} 

\maketitle

\section{Introduction}\label{sec-intro}
Quantum processors rely on classical control and acquisition systems to prepare, manipulate, and read out their quantum states. Whereas open-loop systems execute pre-defined sequences of control operations, closed-loop systems dynamically update their sequences based on measurement outcomes. Real-time feedback control systems are control systems that actuate the processor faster than the dominant dissipation processes. These systems play an increasingly important role in improving the capability of quantum processors across many platforms, including those based on configurations of atoms and molecules.
Their recent deployment in practical settings has enabled the preparation of defect-free configurations of atoms~\cite{Kim2016, Endres2016, Barredo2016}, the realization of non-local connectivity by displacing atoms~\cite{Bluvstein2022}, and the implementation of mid-circuit measurements and feedforward operations to stabilize quantum gates~\cite{Singh2023} and realize quantum error correction codes~\cite{Bluvstein2024}. These demonstrations have relied on custom solutions that interface multiple components, typically utilizing motherboard-based or FPGA architectures~\cite{Endres2016, Wang2023, Bluvstein2024, Manetsch2024}. However, despite their shared architectural principles and workflows, a comprehensive analysis of their underlying hardware and software, along with a quantitative breakdown of their runtime performance, remains largely unexplored. Furthermore, the rapid advances in the capabilities of GPUs, including their processing speed and low-latency data transfers, call for systematic studies of the opportunities they offer for integration into experimental workflows.
\begin{figure}[t!]
\includegraphics[]{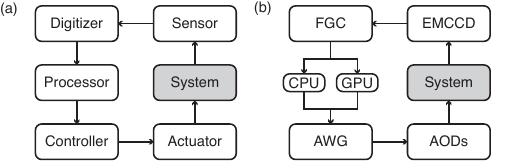}
\caption{
\label{fig-diagram}
\textbf{Diagram of feedback control system.}
(a)~A typical FCS comprises a chain of modules to actuate the state of a physical system based on a sequence of measurements. 
(b)~Our low-latency reconfiguration system is built on the architecture of a typical FCS. An Electron Multiplying Charge-Coupled Device (EMCCD) camera acquires images of configurations of individual atoms. A Frame Grabber Card (FGC) transfers these images to a Central Processing Unit (CPU) or Graphics Processing Unit (GPU). The processor solves an atom reconfiguration problem to compute the sequence of control signals necessary to reach the desired configuration of atoms. An Arbitrary Waveform Generator (AWG) streams these signals to actuate a pair of Acousto-Optic Deflectors (AODs). The resulting multiplexed laser beams update the configuration of atoms. The LLRS repeats the process until the desired state is reached, or the the criteria for solving the problem is no longer met.
}
\end{figure}

In this work, we introduce and characterize an open-source low-latency reconfiguration system (LLRS) for quantum processors based on atoms and molecules. The LLRS features a simple, cost-effective, and extendable motherboard-based architecture that ensures easy deployment and cross-compatibility with multiple devices, including GPUs. Its use of off-the-shelf components provides flexibility in interfacing with diverse devices and enabling new capabilities across various applications. Focusing on atom reconfiguration problems~\cite{Cimring2023, ElSabeh2023, Cooper2024, Afiouni2025}, we describe the system’s architecture and workflow~(Sec.~\ref{sec-system-overview}), followed by a characterization of its runtime performance~(Sec.~\ref{sec-benchmarking}). Our goal is to establish blueprints and benchmarks to expedite the development of similar systems, as well as to identify limitations and opportunities for further improvement~(Sec.~\ref{sec-conclusions}).


\section{System architecture and workflow}\label{sec-system-overview}

We first consider the typical architecture of a feedback control system used to actuate a physical system. The feedback control system comprises a chain of modules operating in a closed loop~(Fig.~\ref{fig-diagram}a). A sensor collects data about the physical system, and a digitizer gathers and distributes the digitized data to a processor. The processor analyzes the data to determine the state of the system, compares the current state against the target state, and solves a combinatorial optimization problem to find the sequence of control operations required to bring the system toward the desired state. The controller translates these control operations into physical signals, either by retrieving them from a database or synthesizing them in real time. These signals are streamed to drive the actuator, which in turn updates the state of the physical system. This process continues in a loop until interrupted or terminated.

Our low-latency reconfiguration system (LLRS) is a specific realization of this architecture specifically designed for actuating fluorescent particles imaged on a camera~(Fig.~\ref{fig-diagram}b). An Electron Multiplying Charge-Coupled Device (EMCCD, Andor iXon Ultra 888) camera images a configuration of atoms or molecules by collecting their scattered photons through an optical microscopy system. 
A Frame Grabber Card (FGC, Active Silicon Firebird 1xCLD-2PE4L) receives the digitized photo-electron counts measured for each pixel of the camera within a pre-defined region of interest. These counts are transferred to the memory of the processor, which we choose as either a CPU (AMD Ryzen Threadripper 2950X) or a GPU (NVIDIA Quadro RTX 4000). The processor analyzes the images to detect the presence or absence of an atom in each trap and infers the configuration of atoms. It then solves an atom reconfiguration problem, determining the sequence of displacement operations required to achieve the desired configuration of atoms. These control operations are translated into polychromatic radio-frequency waveforms, which are then uploaded to the on-board memory of an Arbitrary Waveform Generator (AWG, Spectrum M4i.6622-x8), acting as the controller. The AWG streams the polychromatic waveforms, which are then amplified to drive a pair of Acousto-Optic Deflectors (AODs, AA Opto Electronic DTSX-400) acting as the actuators. These AODs are active optical diffractive devices that deflect a single laser beam at an angle proportional to the frequency of its monochromatic driving field, or multiplex it into multiple laser beams when driven by a polychromatic waveform. Each of the multiplexed laser beams can extract, displace, and implant individual atoms from their traps, effectively updating the configuration of atoms. The reconfiguration process continues for multiple reconfiguration cycles until the desired configuration is reached or the reconfiguration problem is no longer solvable~\cite{Cimring2023}. 


\begin{figure}[t!]
\includegraphics[]{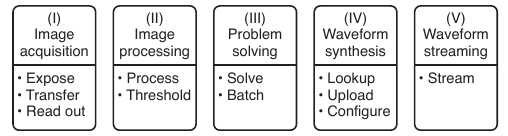}
\caption{
\label{fig-software}
\textbf{Overview of the software architecture.}
Image acquisition is performed by the EMCCD and FGC. Image processing, problem solving, and waveform synthesis are performed by the CPU or GPU. Waveform synthesis and streaming are performed by the processor and AWG. 
}
\end{figure}

A key challenge in designing the LLRS was achieving low-latency operation while maintaining modularity and flexibility across different hardware platforms. Our approach optimizes data transfer pathways and parallel processing strategies to minimize bottlenecks, enabling real-time feedback control at a speed necessary for practical quantum applications. The workflow of the LLRS comprises five modules (Fig.~\ref{fig-software}): (1)~image acquisition, (2)~image processing, (3)~problem solving, (4)~waveform synthesis, and (5)~waveform streaming. We now describe the implementation of each module.



\subsection{Module 1 – Image acquisition} 
The image acquisition module relies on the EMCCD, FGC, and AWG devices. Upon being externally triggered by some external controller, the AWG triggers the EMCCD to open its mechanical shutter and initiate exposure. The photons incident on each pixel of the camera are converted into primary photoelectrons that are stored as charges on a capacitor. After a fixed user-defined exposure time, the EMCCD vertically shifts the photoelectrons to a storage region and then horizontally shifted them to an amplification region that converts them into secondary photoelectrons. These secondary photoelectrons are then digitized by an analog-to-digital converter. As they are collected, the digitized counts are continuously transferred to the FGC over a 3-tap Camera Link interface, which supports higher data transfer rates ($2~\text{GBps}$) than the USB 3.0 protocol supported by the camera (0.625~\text{GBps})~\cite{AndorCameraLink, AndorUSB}. 
After all digitized counts have been received, the FGC transfers them onto the memory of the processor, either to the CPU via Direct Memory Access (DMA) or to the GPU via Remote Direct Memory Access (RDMA). The RDMA protocol transfers the data directly to the onboard memory of the GPU without involving the operating system, thereby speeding up the transfer and reducing timing jitter. This approach eliminates the latency associated with CPU-mediated transfers, which would otherwise introduce bottlenecks and degrade real-time performance. 

\subsection{Module 2 – Image processing} 
The image processing module converts the list of digitized counts into a list of binary occupation numbers, indicating whether each trap is occupied or not. To extract the occupation number of each trap, the module first computes the mean count intensity over a finite box region using a weighted mask chosen as the used-defined normalized point-spread function specified for each trap. This point-spread function has a discretized Gaussian-like profile. The module then solves a binary classification problem using the thresholding method, assigning the presence of an atom if the mean count intensity exceeds the threshold value.
The threshold is chosen to maximize the discrimination probability between the presence and absence of an atom by minimizing the probability of erroneous assignments~\cite{Cooper2024}. This module can be readily extended to tasks such as quantum state estimation and correlation extraction, which are essential for implementing adaptive protocols.
 
\subsection{Module 3 – Problem solving} 

The problem-solving module solves an atom reconfiguration problem using an efficient implementation of an atom reconfiguration algorithm, including red-rec~\cite{Cimring2023, Afiouni2025}, aro~\cite{ElSabeh2023,Afiouni2025}, or bird~\cite{Afiouni2025}. The algorithm returns a partially ordered list of elementary displacement operations which are represented by source and destination indices for each move. The batching routine~\cite{Afiouni2025} translates this list into an ordered list of batched displacement operations, where the operations within each batch are executed simultaneously. Each translated batched operation is associated with a pre-computed waveform stored in a lookup table on the memory of the processor (CPU or GPU). 

\subsection{Module 4 – Waveform synthesis} 

The waveform synthesis module translates the sequence of control operations into a sequence of waveforms that can be streamed on the AWG. Real-time synthesis of these waveforms requires evaluating and summing trigonometric functions, a process that is time-consuming unless specific optimizations are employed. 

One approach is to sum elementary waveforms fetched from a pre-computed database. Another approach is to synthesize the waveform in the Fourier domain and then convert it back to the temporal domain using the Fast Fourier Transform. Here, we use a look-up table that stores precomputed waveforms for each batched control operation from the problem-solving module. This approach minimizes computational overhead and enables efficient reuse of waveforms when working under conditions where they remain unchanged.

The waveform synthesis module fetches the digitized waveforms using the table keys associated with a list of batched control operations. These digitized waveforms are then copied to local memory buffers for subsequent uploading to the onboard memory of the AWG.

Each normalized, discretized waveform, $\tilde{y}(t)=y(t)/\max{(y(t))}$, is evaluated over the finite interval $[0,T]$ with a step size of $\Delta t=1/f_s$, where $f_s$ is the sampling rate of the AWG. The duration of the waveform is set to achieve a desired frequency resolution $\Delta f$, e.g., $T=10~\mu\text{s}$ for $\Delta f = 100~\text{kHz}$.
Each waveform is computed as the normalized sum of elementary discretized waveforms, $y(t) = \sum_j y_j(t)$, where each elementary waveform is a sinusoidal function, $y_j(t) = \alpha_j \sin(2\pi \nu_j t + \phi_j)$, defined by its amplitude $\alpha_j$, frequency $\nu_j$, and phase $\phi_j$. When operating the AWG on multiple channels, the discretized waveforms (1D arrays of samples) for each channel must be interleaved into a single multi-channel waveform. The frequency determines the position of traps, whereas the amplitude determines the trap depth. The amplitudes and phases are typically optimized to minimize spatial inhomogeneity in trapping parameters while maximizing diffraction efficiency~\cite{Endres2016}. 

The lookup table contains both static and dynamic waveforms, depending on whether the parameters are time-dependent or not. Static waveforms hold atoms in place, while dynamic waveforms perform transfer and displacement operations. Dynamic waveforms are defined by time-dependent parameters, $y_j(t) = \alpha_j(t) \sin(\phi_j(t))$, where $d\phi_j(t)/dt=\nu_j(t)$ is the instantaneous frequency. Minimizing atom loss during dynamic operations requires smoothly connecting parameters between neighboring frequency tones. A typical approach to ensuring a smooth connection involves nulling the first few derivatives of the amplitude function at the waveform boundaries and ensuring continuous phase connection, e.g., by choosing 
$\alpha'_j(0)=\alpha''_j(0)=0$, $\alpha'_j(T)=\alpha''_j(T)=0$
$\phi_j(T) = \phi_j(0)$ modulo $2\pi$, $\phi_j'(0) = 2\pi \nu_j$, $\phi_j'(T) = 2\pi \nu_{j+1}$, and $\phi_j''(0) = \phi_j''(T) = 0$.
While some waveforms, cannot satisfy all the conditions above simultaneously, the equality signs can be relaxed by allowing some error tolerance. Our system includes implementations of Tanh, cubic spline, and Erf as transition functions, with the choice determined by experimental optimization.

The size of the lookup table stored on the processor memory depends on the dimensions and the size of problem. For example, when solving 1D reconfiguration problems and restricting displacement operations to contiguous blocks of traps moving forward and backward by a single step, the total number of continuous block displacement waveforms is $N_{t_x}(N_{t_x}-1) \sim \mathcal{O}(N_{t_x}^2)$. Furthermore, when solving 2D problems on grids using red-rec, the size of the lookup table scales as $\mathcal{O}(N_{t_x} (N_{t_y})^2)$. This scaling can be understood by considering that the number of columns is $N_{t_x}$, and the number of block displacement waveforms necessary to solve each column during the reconfiguration step is $N_{t_y}(N_{t_y}-1)$. For the redistribution steps performing displacement operations along rows, the number of interleaved displacement waveforms is $2 N_{t_y} (N_{t_x}-1)$, given $N_{t_y}$ rows and $2(N_{t_x}-1)$  elementary displacement waveforms per row. These waveforms are supplemented by an additional $N_{t_y}(N_{t_y}-1)$ transfer waveforms for extracting and implanting blocks within each column.


\subsection{Module 5 – Waveform streaming}\label{sec-wfm-streaming-overview}
The waveform streaming module relies on the AWG playing waveforms and distributing them to the AODs through a chain of analog filters and high-power amplifiers. The AWG can be operated in either first in, first out (FIFO) mode or sequence mode. In FIFO mode, the AWG streams waveforms from a memory buffer that is continuously populated by the processor. To avoid buffer underrun errors, the processor must upload waveforms faster than they are streamed; however, due to unpredictable timing jitter in waveform generation and data transfer, the relevant time scale is not the mean upload time but the worst-case upload time. This constraint amplifies the trade-off between speed (latency) and robustness (throughput). Optimizing the system for robustness requires choosing a large buffer, which increases the minimum achievable latency. Although our system is specified for FIFO-mode operation at the maximum sampling rate, we have had limited success in operating in this mode due to buffer underrun errors resulting from timing jitter from the processor. Consequently, we choose to operate in sequence mode. 

In sequence mode, the AWG streams a sequence of waveforms saved in the data memory of the AWG based on a list of instructions encoded in the sequence memory. The data memory is a physical memory buffer of size $2^{32}$ bytes, which can contain up to $2^{31}$ samples with 2 bytes per sample. For a sample rate of $6.24 \cdot 10^6$ samples per second, the buffer can store more than 6 seconds of streaming time. This buffer can be partitioned into a maximum of 4096 segments. We choose the size of each segment to contain a buffer of $32$ waveforms. This size is chosen to minimize the possibility of underrun based on the waveform uploading runtime~(see Fig.~\ref{fig-wfm-streaming} and Sec.~\ref{sec:wfm_streaming}). 

A key feature of the AWG is its ability to upload data to all buffer segments during streaming, except for the one currently playing. The sequence memory consists of 4,096 step registers, each programmable to play a specific data segment and jump to the next step based on user-defined conditions. This sequence memory can be dynamically updated while the AWG is streaming waveforms. However, if the currently playing step is updated, the changes will not take effect immediately; instead, they will be applied after a non-deterministic number ($geq 1$) of playbacks of the step.

\begin{figure}[t!]
\includegraphics[]{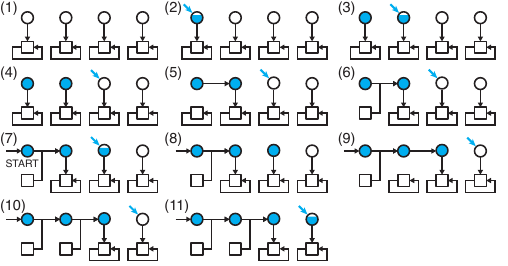}
\caption{
\label{fig-sequence-memory}
\textbf{Sequence memory.}
The sequence memory of the AWG is segmented into an idle step (not shown), a sequence of control steps (circles), and a sequence of failsafe steps (squares). Each control step points to a failsafe step, which loops back to itself, ensuring atoms remain trapped in case of a timing underrun error. Waveform segments (blue disks) associated with each control step are continuously uploaded to the data memory. Directional pointers are updated when the waveform segments associated wit the next control step has been filled. The reconfiguration cycle starts when the first two control steps have been filled by triggering a jump from the idle step to the first control step.
}
\end{figure}    

When initializing the AWG, we partition the data memory into multiple segments, each associated with a specific waveform type. First, we define an idle segment containing a static waveform that keeps all traps on to retain the atoms. Second, we define a failsafe segment with a similar static waveform to maintain the atoms in their traps in the event of a timing underrun error. Third, we allocate multiple equally sized control segments to host the dynamic waveforms responsible for transfer and displacement operations. These segments are dynamically updated throughout the reconfiguration cycle. Finally, we define a double-sized control segment to store all necessary waveforms for reconfiguration when the total number of waveforms is less than twice the segment size.

We partition the sequence memory into an idle step, control steps, and failsafe steps (see Fig.~\ref{fig-sequence-memory}). Each control step is paired with a corresponding failsafe step to detect and identify the occurrence of a timing underrun error. The idle step is configured to play the idle segment and continuously loops back to itself. Each control step, including the double-sized control step, is associated with its respective control segment and points to an individual failsafe step. Each failsafe step plays the failsafe segment and loops back to itself, ensuring atoms remain trapped in case of a timing underrun error. After each reconfiguration cycle, the sequence memory is reset to its initial state while performing image acquisition.

To initiate the reconfiguration cycle, we first pre-upload two segments to the data memory of the AWG (2-4). Next, we update the sequence memory: one update directs the previous control step to the next control step (5), while another directs the previous failsafe step to the next control step (6). Streaming begins by updating the idle step to point to the first control step, while additional segments are concurrently uploaded to fill the next control steps (7). Pointers are continuously updated after each segment upload (8-11). This process continues until all control segments have been uploaded, at which point the last control step is directed back to the idle step, restoring idle status at the end of the reconfiguration cycle. The minimum achievable latency is thus the time required to upload the first two segments, plus the sequence memory updates necessary to transition the AWG out of the idle state.

\section{Runtime performance}\label{sec-benchmarking}
\begin{table*}
\begin{center}
        \begin{tabular}{|l|p{35mm}|p{35mm}|p{35mm}|} 
            \hline
               & \multicolumn{3}{c|}{Reconfiguration cycle time~(ms)} \\ 
            \hline
            & \multicolumn{1}{c|}{Exact 1D} & \multicolumn{1}{c|}{red-rec} & \multicolumn{1}{c|}{red-rec} \\ 
                & \multicolumn{1}{c|}{$N_a = 1 \times 32$} & \multicolumn{1}{c|}{$N_a = 16 \times 16$} & \multicolumn{1}{c|}{$N_a = 32 \times 32$} \\ 
                & \multicolumn{1}{c|}{$N_t = 1 \times 64$} & \multicolumn{1}{c|}{$N_t = 16 \times 32$} & \multicolumn{1}{c|}{$N_t = 32 \times 64$} \\ 
                & \multicolumn{1}{c|}{ROI~=~$16\times1024$} & \multicolumn{1}{c|}{ROI~=~$256\times512$} & \multicolumn{1}{c|}{ROI~=~$512\times1024$} \\ 
            \hline
            \textbf{I. Image acquisition} & $25.454\pm0.695$ & $30.850\pm0.588$ & $46.250\pm0.707$ \\ 
            \hline
            \hspace{3mm} \small I.1 Exposure* &\small20.000 & \small20.000 & \small20.000 \\ 
            \hspace{3mm} \small I.2 Vertical frame transfer* ($4.33~\mu s$ / px) &\small4.499 &\small4.499 &\small4.499\\ 
            \hspace{3mm} \small I.3 Frame readouts* (30~Mpxps) &\small0.677&\small5.923&\small20.549\\ 
            \hspace{3mm} \small I.4 DMA data transfer &\small0.278 &\small0.428 &\small1.202\\ 
            \hline
            \textbf{II. Image processing} & $0.259\pm0.108$ & $0.281\pm0.158$ & $0.277\pm0.084$ \\ 
			\hline
            \hspace{3mm} \small II.1 Deconvolution &\small0.259  &\small0.279 &\small0.269 \\ 
            \hspace{3mm} \small II.2 Thresholding  &\small0.001 &\small0.002&\small0.007\\ 
            \hline
            \textbf{III. Problem solving} & $0.008\pm0.001 $ & $0.059\pm0.005$ & $0.177\pm0.014$ \\ 
            \hline
            \hspace{3mm} \small III.1 Solving&\small0.005  &\small0.035 &\small0.100 \\ 
            \hspace{3mm} \small III.2 Batching &\small0.003 &\small0.024&\small0.077\\ 
            \hline
            \textbf{IV. Waveform synthesis} &$0.630\pm0.268$ & $ 0.663\pm0.117$ & $ 0.635\pm0.247$ \\ 
            \hline
            \hspace{3mm} \small IV.1 Waveform lookups ($2\times32=64~\text{wfms}$)  &\small0.078 &\small0.046&\small0.038\\ 
            \hspace{3mm} \small IV.2 Waveform uploads (2 segments) &\small0.553 &\small0.593&\small0.574\\ 
            \hspace{3mm} \small IV.3 Memory configuration &\small0.000 &\small0.025&\small0.023\\ 
            \hline
            \textbf{V. Waveform streaming} &$0.746\pm0.004$ & $2.765\pm0.392$ & $8.496\pm0.577$ \\ 
            \hline
            \hspace{3mm} \small V.1 Streaming* &\small$0.640 \pm 0.000$ &\small$2.667 \pm 0.355$&\small$8.392 \pm 0.533$\\ 
            \hspace{3mm} \small V.2 Other &\small$0.106\pm0.004$ &\small$0.098\pm0.191$&\small$0.104\pm0.231$\\ 
            \hline
            \textbf{Total} & $27.097\pm1.076$ & $34.618\pm1.260$ & $55.835\pm1.629$ \\ 
            \hline
            \hline
        \end{tabular}
\end{center}

\caption{Tabulated runtime values for solving reconfiguration problems on chains and grids. Values are averaged over one thousand realizations of a typical problem. The first column reports values for preparing center-compact configurations on chains using the exact 1D algorithm. The second column report values for preparing center-compact configurations on grids using the latest version of the red rec CPU algorithm. Values reported for steps labeled by stars are theoretical values. The problem instances are randomly generated based on an $\epsilon = 0.6$ initial probability of the presence of a particle in a trap.}\label{table:runtime} 
\end{table*}

\begin{figure}[t!]
\includegraphics[]{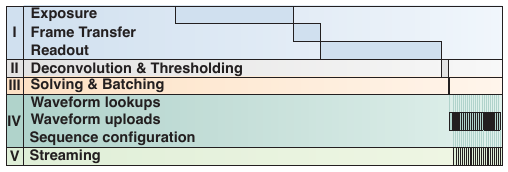}
\caption{
\label{fig-timing}
\textbf{Runtime.}
A graphical representation of the typical runtime for executing each of the five modules in the preparation of a $32\times32$ atom configuration using red-rec. The primary timing bottleneck is image acquisition. Waveform uploads occur concurrently with waveform streaming.
}
\end{figure}    

To characterize the runtime performance of our system, we measure the computational runtime for each step of a typical reconfiguration cycle (Table~\ref{table:runtime}, Fig.~\ref{fig-sequence-memory}). While unit tests are performed out of loop, measurement times are taken in-loop; we have confirmed that the in-loop results are consistent with the out-of-loop results.
The measured values are compared against their theoretical estimates whenever possible.

We specify a reconfiguration problem by the geometry and size of the static trap array, as well as the number of atoms in the desired center-compact configuration. A problem instance is defined by randomly sampling $N_a^T$ atoms in a trap array of $N_t=N_{t_x}\times N_{t_y}$ traps. We typically choose  $N_{t_x}=\sqrt{N_a^T}$ and $N_{t_y}=2 N_{t_x}$, as they achieve a baseline success probability of $\bar{p}=0.5$ in the absence of loss when the loading efficiency is $\epsilon=0.5$.

We perform our runtime benchmarking analysis using synthetic images. When triggered by the AWG, the EMCCD camera acquires an image with its mechanical shutter closed. Since the image consists only of background noise, the image processing module extracts a binary array of zeros. This array is then replaced by the expected occupation state, which was computed by pre-solving the atom reconfiguration problem in an out-of-loop step. Pre-solving the various problem instances in this way avoids the need to execute a random number generator during runtime benchmarking, thereby preventing any disturbance to the processor. After solving a thousand randomly sampled problem instances, we compute the mean and standard deviation of the runtime distribution obtained for each step and module of the reconfiguration process. 

The runtime values are reported in Table~\ref{table:runtime} for three typical reconfiguration problems.  Given that the aro algorithm is too slow for real-time operations~\cite{Afiouni2025} and that the runtime of bird is comparable to the one of red-rec~\cite{Afiouni2025}, we focus on benchmarking runtime performance for red-rec. The first problem is preparing a center-compact chain of atoms. The two other problems are preparing a center-compact configuration of atoms in a rectangular grid of atoms using an improved version of the red-rec algorithm~\cite{Cimring2023, Afiouni2025}. We now analyze these results for each step of the reconfiguration cycle.


\subsection{Module 1 – Image acquisition}

The image acquisition module involves four key steps. First, the sensor is exposed for a predefined exposure time following an external trigger. Second, the primary photoelectrons are vertically transferred from the exposure region to the storage region. Third, the accumulated charges in each pixel are sequentially amplified and digitized. Fourth, the digitized counts are transferred from the camera to the processor via the FGC.

The exposure time is set by the user to unambiguously distinguish between the presence and absence of an atom in a single-shot image. This setting balances the number of photons collected from individual atoms against background noise. The photon count from atoms is influenced by the scattering rate of the atoms, the collection efficiency of the optical imaging system, and the detection efficiency of the camera~\cite{Sortais2007} . The scattering rate, in turn, is proportional to the intensity of the near-resonant excitation light, which, in lossless imaging scenarios, must be carefully managed to avoid ejecting atoms from their traps due to heating. Background photon contributions stem from cosmic rays, charge-induced counts, which are affected by charge-shifting speed, as well as stray light sources and unfiltered cooling, trapping, and imaging light. The discrimination threshold also depends on the gain and readout noise of the analog-to-digital converter. Here, we set the exposure time to a typical lossless imaging value of $20~\text{ms}$. 

\begin{figure}[t!]
\includegraphics[]{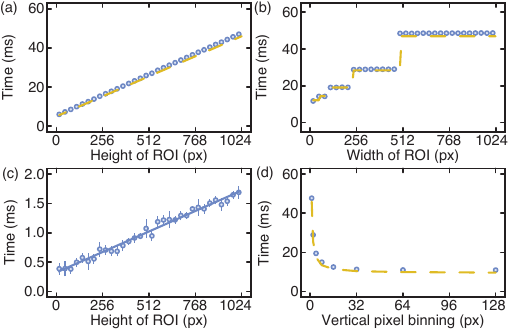}
\caption{
\label{fig-image-acquisition}
\textbf{Image acquisition.}
(a)~Frame transfer and readout time of a full-width image of various ROI heights (blue disks), excluding the exposure time, from EMCCD to CPU memory via the FGC. The measured values (blue disks) differ from the linear-scaling prediction (yellow dashed line) due to the finite transfer time from the FGC to the CPU memory.
(b)~Readout and transfer time of a full-height image of various ROI widths (blue disks), excluding the exposure time, with predicted values (yellow line). The steps correspond to powers of two of the width.
(c)~Difference between the measured and predicted times (blue circles). The transfer rate is estimated to be 1.43~GBps from a linear fit (blue line).
(d)~Readout and transfer time of a full-size image with vertical binning (blue disks). The predicted value (yellow line) follows an inverse power law of two. Horizontal binning does not change the acquisition time. 
}
\end{figure}    


The vertical transfer and readout times of the camera frame depend on the characteristics of the camera sensor~\cite{CCDDataSheet, Schwegler2018, Wang2023}. Our EMCCD has a CCD201-20 frame transfer, electron multiplying CCD sensor with an active exposure area of $1024\times1024~\text{px}$ and a pixel size of $13\times13~\mu\text{m}$. The total sensor has dimensions of $1056 \times 2069$ pixels in the horizontal and vertical directions, respectively. In the vertical direction, the 2069 pixels are partitioned from top to bottom into 5 dark reference rows, 1 transition row, an image section of 1024 active rows, 2 transition rows, and a storage section of 1037 rows. The reference rows, transition rows, and storage section are shielded from external light, exposing only the image section. In the horizontal direction, the sensor has 1056 pixels, but the image section is shielded on both sides by 16 dark reference pixels, providing an effective image area of $1024\times1024~\text{px}$. The sensor is located above a single-row amplification and readout section containing 1056 registers which is next to into a distribution chain of 468 pixels and a multiplication chain of 604 pixels. The readout section is enclosed by 16 additional overscan pixels on each side before the linkage to the analog-to-digital converter. 

We estimate the vertical frame transfer time from the image section to the storage section to be $(1037+2)/v_V=4.499~\text{ms}$, where $v_V=4.33~\mu\text{s/px}$ is the vertical shift speed per row. Assuming that the bottom row of the storage section is shifted to the readout region as soon as the receiving registers have been emptied, the frame readout time for each row is equal to the vertical shift time, $t_V=1/v_V$, plus the latency time for reading out the stray charges by horizontally shifting the pixels along the readout region ($N_H$ columns of the image), the dark references (32 pixels), and the overscan elements (16 pixels).  Given $N_V$ rows, the frame readout time is thus $N_V (t_V+(N_H + 48)/v_H)$, which is equal to $41~\mu\text{s}$ at full frame for $N_V=1024, N_H=1024$, and horizontal shift rate of $v_H=30~\mu\text{s/px}$. There is an additional constant offset for horizontally shifting the pixels through the dump region which includes the chain of 1072 gain elements and the overscan elements (16 pixels) at the start of the readout. 

The frame transfer and readout time can be reduced by restricting the region of interest (ROI), defined as the contiguous set of pixels starting with the one closest to the detection region. As expected, our measurements shows that the frame transfer and readout time depends linearly on the height of the ROI~(see Fig.~\ref{fig-image-acquisition}a). This time can be further reduced by restricting the width of the region of interest to a power of two~(see Fig.~\ref{fig-image-acquisition}b).
The readout time can be further reduced by vertically binning the pixels, which decreases by a factor of two with each doubling of the binning size in powers of two (Fig.~\ref{fig-image-acquisition}d). However, horizontal binning does not affect the readout time.

We explain the difference between the measured and predicted readout time by the transfer time from the FGC to the processor. A linear fit to the residuals~(see Fig.~\ref{fig-image-acquisition}c) show a transfer rate of $1.431~\text{GBps}$ for $2~\text{B/px}$, 
which is consistent to the rate of $1.7~\text{GBps}$ expected for the FGC using DMA transfer via 4-lane Gen2 PCIe protocol. A similar transfer rate of $1.206~\text{GBps}$ is obtained when transferring data from the FGC to the GPU using RDMA transfer.

\subsection{Module 2 – Image processing and analysis}

\begin{figure}[t!]
\includegraphics[]{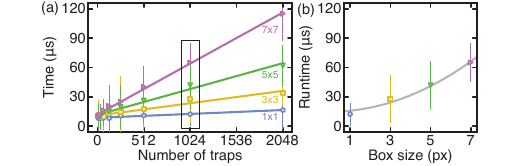}
\caption{
\label{fig-image-processing}
\textbf{Image processing.}
(a)~Processing time for extracting the occupation state of a trap array of various sizes from a full-size image using the weighted filtering method. The time increases linearly with the number of traps and (b)~quadratically with the side length of the squared-boxed filter. The discretized point-spread function is chosen as the weighted filter defined on a box of sizes $1 \times 1$ (blue disks), $3 \times 3$ (yellow squares), $5 \times 5$ (green inverted triangles), and $7 \times 7$ (purple right-rotated triangles) pixels. 
}
\end{figure}

The image processing module involves two steps. First, the mean weighted intensity at the location of each trap is calculated by deconvolving the image with a kernel defined by the point-spread function (PSF) associated with each trap. Second, the counts are thresholded to infer the presence or absence of an atom in each trap. The image processing time grows linearly with the number of traps~(see Fig.~\ref{fig-image-processing}a) and quadratically with the PSF box width~(see Fig.~\ref{fig-image-processing}b). Our current implementation is based on OpenMP, an open standard for parallel programming in shared-memory architectures. It exploits a multi-threaded method by parallelizing computations across distinct boxes, minimizing shared memory overhead. Under optimal conditions for initializing threads, this implementation achieves a deconvolution time of $64~\mu\text{s}$ for a configuration of $N_t=32\times32=1024$ traps using a PSF defined over a box containing $7\times 7=49~\text{px}$. However, our benchmarks show that the initialization overhead significantly increases during the reconfiguration process (Table~\ref{table:runtime}). 
 
Thresholding is a simple comparative operation that is nearly instantaneous as a result of our processor's single instruction multiple data (SIMD) capabilities.


\subsection{Module 3 – Problem solving}

We solve reconfiguration problems on chains using an improved implementation of the exact 1D algorithm~\cite{ElSabeh2023, Afiouni2025}, and on rectangular grids using an improved version and implementation of the red-rec algorithm~\cite{Cimring2023, Afiouni2025}. These improved implementations outperform previous ones in both runtime and operational performance, approaching the efficiency of the aro algorithm~\cite{ElSabeh2023}, which has also been sped up but not to a level suitable for real-time operations. 

\begin{figure}[t!]
\includegraphics[]{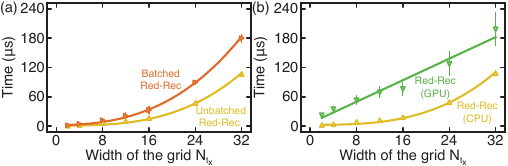}
\caption{
\label{fig-problem-solving}
\textbf{Problem solving.}
(a)~Processing times for preparing a center-compact configuration of $N_{t_x}^2$ atoms in a grid of $N_{t_x}\times N_{t_y}$ traps with $N_{t_y}=N_{t_x}/0.6$ using red-rec with (orange triangles) and without (yellow triangles) batching. The unbatched red-rec implementation scales as $N_t^{3/2}\sim N_{t_x}^3$ (yellow line), whereas the batched one includes an additional quadratic factor.
(b)~Processing time for solving the same reconfiguration problems using the unbatched red-rec algorithm implemented on a CPU or a GPU.
}
\end{figure}

The measured runtime of red-rec without batching scales as $\mathcal{O}(N_{t_x}^3)$ (see yellow line in Fig.~\ref{fig-problem-solving}a,~b). Batching scales as $\mathcal{O}(n)$ where $n$ is the solution length. Considering a linear scaling of the solution length with respect to the problem size, the additional batching runtime can be upper bounded by a factor of $\mathcal{O}(N_{t_x}^2)$ (see orange line in Fig.~\ref{fig-problem-solving}a). When preparing a configuration of $N_a^T=32\times32$ atoms, the measured runtime approaches $106(6)~\mu\text{s}$ and $180(9)~\mu\text{s}$ without and with batching, respectively. To justify the use of CPU, we compare these results against those obtained for a parallel implementation of red-rec on the GPU (see green line in Fig.~\ref{fig-problem-solving}b). The runtime for the GPU implementation exhibits a finite initialization time of $7~\mu\text{s}$ and approximately linear scaling with the nearest power of two of $N_{t_x}$. The use of a GPU might be justified for preparing configurations containing more than $N_{t}=42\times42=1764$ atoms.

\subsection{Module 4 – Waveform synthesis}
Waveform synthesis is done concurrently with waveform streaming. The only runtime costs contributing to latency are the time needed to look up and upload two full segments of 32 waveforms from the lookup table and update them in the sequence memory of the AWG. These two segments are necessary to avoid underrun errors.

Our waveform table implementation uses contiguous buffers, allowing access to a specific waveform by simply calculating the index. As a result, lookups are done in constant time regardless of the database size. The total runtime grows linearly with the number of waveforms, which equals the number of batched control operations multiplied by the time it takes to access and copy each waveform.

Our benchmarking results indicate that the lookup time per waveform is $0.21~\mu\text{s}$, corresponding to $14~\mu\text{s}$ for 2 segments of 32 waveforms. The upload time is $252~\mu\text{s}$ per segment, for a total of $506~\mu\text{s}$. Updating the sequence memory takes $4~\mu\text{s}$. However, the delay time before the sequence memory update is in effect is non-deterministic due to internal data processing performed by the proprietary hardware of the AWG.


\subsection{Module 5 – Waveform streaming}\label{sec:wfm_streaming}

\begin{figure}[t!]
\includegraphics[]{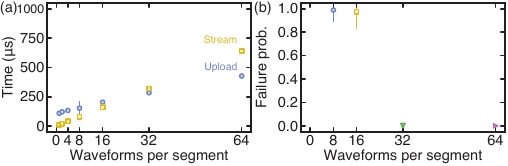}
\caption{
\label{fig-wfm-streaming}
\textbf{Waveform streaming.}
(a)~Measured time for uploading a waveform segment of various lengths to the AWG memory (blue disks). The upload time is less than the predicted waveform stream time (yellow squares) for segments containing more than 32 waveforms per segment.
(b)~Probability of encountering an underrun error as the number of waveforms per segment increases. The system is stable when the stream time is larger than the upload time. A small failure probability remains due to unexpected timing events.
}
\end{figure}

Waveform streaming occurs uninterrupted until all waveforms have been played. The total runtime is equal to the number of waveforms multiplied by the duration of each waveform. As noted previously (see Sec.~\ref{sec-wfm-streaming-overview}), waveform streaming is performed concurrently with waveform synthesis; the waveforms are continuously uploaded to the data memory of the AWG, while the sequence memory is updated to ensure smooth transitions between the various segments. Continuous streaming without underrun errors requires the upload time of each segment to be shorter than the stream time (see Fig.~\ref{fig-wfm-streaming}a,~b). This condition is met when more than 32 waveforms are uploaded per segment, reducing the probability of failure to less than one in every $257$ reconfiguration events (see Fig.~\ref{fig-wfm-streaming}b). Underrun errors occur due to timing jitter from non-deterministic execution, possibly induced by the operating system. When an underrun error happens, the AWG switches to a failsafe sequence that plays a static waveform, holding all atoms in their trap. The worst outcome is the loss of all extracted atoms, resulting in a reduction of the probability of solving the problem, and thus a reduction in the operational performance.



\section{Conclusions}\label{sec-conclusions}
We presented the architecture and workflow of a low-latency feedback system and quantified its runtime performance for solving atom reconfiguration problems on chains and grids. The associated software is publicly available as an open-source package, enabling cross-comparison across different systems and implementations. The system is a low-cost, motherboard-based system that comprises a frame grabber card, a processor, and an arbitrary waveform generator. The system supports the remote direct memory access (RDMA) protocol to reduce the transfer time between the GPU and other peripherals, as well as lookup tables to reduce the synthesis time of control waveforms. The system is compatible with parallel processing on GPUs, although its use is only justified when solving large reconfiguration problems. 

The system realizes a simple five-step workflow that acquires and processes images from a camera, solves combinatorial optimization problems on a processor, and synthesizes and stream control waveforms to actuate a configuration of atoms. The system achieves high runtime efficiency by leveraging highly optimized implementations of reconfiguration algorithms~\cite{Afiouni2025}, concurrent processes, and look-up tables.

Our runtime benchmarking results show that the latency is limited by the image acquisition and waveform streaming modules. The image acquisition time can be reduced by cropping the region of interest, binning pixels, increasing vertical shift speed and readout rate, albeit at cost of a potential decrease in discrimination fidelity. The frame transfer and readout time may be reduced by implementing advanced pixel-shifting algorithms, or using a different camera, either an EMCCD camera with a smaller sensor or having multiple analog-to-digital converters. Another solution is using different sensing technologies, such as a CMOS camera operating in global shutter mode or SPAD arrays, which allows measuring pixels in parallel. The streaming time is ultimately limited by the efficiency of the reconfiguration algorithms, as well as the duration of each waveform. Further reducing waveform time involves a tradeoff between shorter durations and small probability of atom loss during displacement and transfer operations.

Our system architecture can be readily be adapted to build feedback control systems using different devices for different applications. For example, EMCCD camera can be replaced by a CMOS camera, the processor by a field-programmable gate array (FPGA), and the AOD by a spatial light modulator or digital micromirror device. Moreover, our system can readily be used for applications other than solving reconfiguration problems including optimizing and stabilizing optical intensity and trap depth, applying atom-selective control pulses, implementing variational algorithms and optimal control pulse synthesis, as well as adaptive protocols for metrology and quantum error correction~\cite{Bluvstein2024, Kurman2024, Barber2025}.

The key advantages of this system are its low cost, its flexibility in integrating new devices and applications, and its ease of programming. It is also compatible with GPUs that enable fast matrix multiplications, parallel computing applications, and auto-differentiation. However, it suffers from its dependence on an operating system, which prevents the implementation of deterministic timing, as provided by FPGAs and real-time operating systems.

Our source code is open-source licensed and is available for use in a public repository.

\section{Acknowledgment}
We acknowledge contributions from Sailesh Bechar, Brooke MacKenzie Dolny, Zefei Ou, Jessica Bohm, Zhiqian (Jessie) Ding, Zewen (Wendy) Lu, Laurent Zheng, and Xiang Wen (Evan) Yu. This research was funded thanks in part to CFREF.


%

\end{document}